\documentclass[prr,aps,twocolumn,showpacs]{revtex4-1} 
\usepackage{amsmath,amssymb,amsthm}
\usepackage{graphicx}
\usepackage{subfigure}
\usepackage{amsmath}
\usepackage{amsfonts,amssymb}
\usepackage{bm}
\usepackage{float}
\usepackage{color}
\usepackage{sidecap}
\allowdisplaybreaks
\usepackage{soul}
\setstcolor{red}
\usepackage[normalem]{ulem}
\usepackage{hyperref}
\hypersetup{colorlinks=true,breaklinks,urlcolor=blue,linkcolor=blue,citecolor=blue}

\begin{document}

\title{Winding Numbers and Generalized Mobility Edges in Non-Hermitian Systems}
\author{Qi-Bo Zeng$^{1}$}
\author{Yong Xu$^{1,2}$}
\email{yongxuphy@tsinghua.edu.cn}
\affiliation{$^{1}$Center for Quantum Information, IIIS, Tsinghua University, Beijing 100084, People's Republic of China}
\affiliation{$^{2}$Shanghai Qi Zhi Institute, Shanghai 200030, People's Republic of China}

\begin{abstract}
The Aubry-Andr\'e-Harper (AAH) model with a self-dual symmetry plays an important role in studying the Anderson localization.
Here we find a self-dual symmetry determining the quantum phase transition between extended and localized states
in a non-Hermitian AAH model and show that the eigenenergies of these
states are characterized by two types of winding numbers. By constructing and studying a non-Hermitian generalized AAH model,
we further generalize the notion of the mobility edge, which separates the localized and extended states in the energy spectrum of disordered systems,
to the non-Hermitian case and find that the generalized mobility edge is of a topological nature even in the open boundary geometry in the sense that the energies of localized and extended states exhibit distinct topological structures in the complex energy plane.
Finally, we propose an experimental scheme to realize these models with electric circuits.
\end{abstract}

\maketitle

\section{Introduction}\label{sect1}
Anderson localization~\cite{Anderson1958} is a ubiquitous phenomenon in disordered physical systems. Due to the destructive interference of scattered waves, the states in the system can become localized~\cite{Ramakrishnan1985RMP,Abrahams2010}.
So far, the Anderson localization has been experimentally observed in various platforms, such as light~\cite{Wiersma1997Nat,Scheffold1999Nat,Schwartz2007Nat,Segev2013Nat}, cold atoms~\cite{Billy2008Nat,Roati2008Nat,Luschen2018PRL}, microwave~\cite{Dalichaouch1991Nat,Chabanov2000Nat,Pradhan2000PRL}, and photonic lattices~\cite{Lahini2009PRL}. In three-dimensional systems with uncorrelated (random) disorders, the localization phase transition occurs with a mobility edge, which is defined as the energy separating the extended and localized states in the energy bands~\cite{Mott1987JPC}. Though such phase transition is excluded by scaling theory in lower-dimensional disordered systems~\cite{Abrahams1979PRL}, it still happens in systems with correlated disorders. One paradigmatic example is the one-dimensional (1D) Aubry-Andr\'e-Harper (AAH) model~\cite{Aubry1980,Harper1955}, which is a lattice model with incommensurate onsite modulations and is of great importance in studying the Anderson localization in quasicrystals~\cite{Siggia1983PRL,Kohmoto1983PRL,DasSarma1990PRB,DasSarma2009PRA,DasSarma2010PRL}. Because of the self-dual symmetry in the original AAH model, no mobility edge is observed. Nevertheless, with appropriately designed onsite modulations or long-range hopping, the mobility edge will emerge in these quasiperiodic lattice models~\cite{DasSarma1988PRL,Izrailev1999PRL,Ganeshan2015PRL,Deng2019PRL}.

During the past few years, non-Hermitian topological systems have been extensively studied both theoretically and experimentally~\cite{Rudner2009PRL,Esaki2011PRB,Bardyn2013NJP,Poshakinskiy2014PRL,Zeuner2015PRL,Yuce2015PLA,Malzard2015PRL,Rudner2016arxiv,Aguado2016SR,Lee2016PRL,Molina2016PRL,Joglekar2016PRA,Zeng2016PRA,Weimann2017NatM,Leykam2017PRL,
Xu2017PRL,Menke2017PRB,Xiao2017NatP,Lieu2018PRB,Zyuzin2018PRB,Fan2018PRB,Alvarez2018PRB,HZhou2018Sci,Yin2018PRA,Shen2018PRL,Kunst2018PRL,Xiong2016,Alvarez2018PRB,Xiong2018JPC,Yao2018PRL1,Yao2018PRL2,Gong2018PRX,Kawabata2018PRB,Takata2018PRL,
YChen2018PRB,RYu2018arxiv,Kawabata2019NatC,Yang2019PRB,Song2018,Wang2019PRB,Ueda2019PRL,Edvardsson2019PRB,Xu2019FrontPhy,Herviou2019PRA,HZhou2019PRB,Kunst2019PRB,
Cerjan2019NatP,Jianbin2019PRL,Kou2019PRB,Danwei2019SciCh,Thomale2019arxiv,Xue2019arXiv,Thomale2019arXiv,Chuanwei2019PRL,Fan2019PRB,Sato2019PRX,Jiang2020EPJB,Zhang2019arxiv,Okuma2020PRL,Song2019PRL,Eunwoo2020PRB,Yoshida2019PRR,
Zeng2019arxiv,Slager2020PRL}. The interplays between the topology and non-Hermiticity result in a plethora of exotic phenomena that have no Hermitian counterparts, e.g., the Weyl exceptional ring~\cite{Xu2017PRL}, the anomalous edge mode~\cite{Lee2016PRL}, the point gap~\cite{Gong2018PRX}, and the non-Hermitian skin effect~\cite{Xiong2016,Alvarez2018PRB,Xiong2018JPC,Yao2018PRL1,Yao2018PRL2}. Recently, the topological phases in the non-Hermitian AAH model have been explored~\cite{Zeng2019arxiv}. The Anderson localization phenomena in such non-Hermitian quasiperiodic as well as disordered systems have also been investigated~\cite{Hatano1996PRL,Hatano1998PRB,Zeng2017PRA,Longhi2019PRL,Jiang2019PRB,Longhi2019PRB}. Reference~\cite{Longhi2019PRL} shows that the Anderson localization phase transition in a 1D non-Hermitian quasicrystal with onsite gain/loss is topological and can be characterized by a winding number. But whether the self-dual symmetry exists in non-Hermitian systems remains elusive. Moreover, though the mobility edge has been found in the
disordered Hatano-Nelson model~\cite{Gong2018PRX}, the topological feature does not exist in an
open boundary geometry.
One may wonder whether distinct topological structures can emerge for the energies of
extended and localized states so that a topological mobility edge appears in a system
with open boundaries.
This seems impossible as it has been shown that the energy spectrum
cannot exhibit a nonzero winding number in a system with open boundaries~\cite{Okuma2020PRL,Zhang2019arxiv}. However,
the winding number contributed by the complex onsite potential
has not been considered there.

Recently, electric circuits have been shown to be a powerful platform to simulate various topological
phases, which have been extensively explored both theoretically and experimentally~\cite{RYu2018arxiv,Jiang2019PRB,Zeng2019arxiv,Imhof2018Nat,Thomale2019PRL,Yang2019PRL,Thomale2019arxiv,Xu2019arxiv}. The circuits can be implemented flexibly and the topological features can be extracted by measuring electrical signals, such as the voltages in the system. In Ref.~\cite{Thomale2019arxiv}, the breakdown of bulk-boundary correspondence and the non-Hermitian skin effect have already been observed in a nonreciprocal topolectric circuit. It will be interesting to use similar schemes to study the Anderson localization and mobility edges in non-Hermitian systems.

In this paper, we study the self-dual symmetry, the winding numbers and the mobility edges in
non-Hermitian AAH models.
(i) We show that there are two types of winding numbers: $W_h$ arising from asymmetric hopping and
$W_o$ arising from the complex onsite potential.
(ii) We find a self-dual symmetry in a non-Hermitian AAH model with asymmetric hopping
determining the quantum phase transition between extended and localized states.
The energies of both localized and extended states form loop structures in the complex energy plane that are
characterized by the winding number $W_o$ and $W_h$, respectively,
under periodic boundary conditions (PBCs), and by $W_o$ under open boundary conditions (OBCs).
(iii) We further construct a non-Hermitian generalized AAH model with $\mathcal{PT}$ symmetry
that hosts both localized and extended states in the energy spectra.
By generalizing the mobility edge in a Hermitian system to a non-Hermitian one,
we find that the generalized mobility edges under both PBCs and OBCs
are topological in the sense that the energy spectra of the localized and extended states
exhibit nonzero and zero winding numbers. With weak asymmetric hopping breaking the $\mathcal{PT}$ symmetry, we find that
the energy spectra of both localized and extended states obtained under PBCs
form loop structures characterized by the winding number $W_o$
and $W_h$, respectively. Under OBCs, only the latter spectra exhibit nonzero $W_o$.
We also find loop structures in the energy spectra obtained under OBCs characterized
by $W_o$; a single loop structure consists of the energy spectra of both localized and extended states.
For larger asymmetric hopping, we demonstrate that the energy spectra of extended states can also
possess a winding number (nonzero $W_o$).
(iv) Finally, we propose a practical experimental scheme with electric circuits to simulate these models
and detect the predicted features.

The rest of the paper is organized as follows. In Sec. \ref{sect2}, we introduce the self-dual symmetry in non-Hermitian AAH models. Then we explore the properties of mobility edges in non-Hermitian systems in Sec. \ref{sect3}. Finally we present the experimental proposal for realizing the non-Hermitian models by using electrical circuits in Sec. \ref{sect4}. The last section (Sec. \ref{sect5}) is dedicated to a brief summary.

\section{Self-dual symmetry}\label{sect2}
We start by considering a non-Hermitian AAH model described by
\begin{equation}\label{HrEq}
\hat{H}=\sum_{j}[e^{i\phi_h/L}t_L\hat{c}_j^\dagger \hat{c}_{j+1} + e^{-i\phi_h/L}t_R\hat{c}_{j+1}^\dagger \hat{c}_j +
V_j \hat{c}_j^\dagger \hat{c}_j],
\end{equation}
where $\hat{c}_j$ ($\hat{c}_j^\dagger$) is the annihilation (creation) operator of a spinless particle at site $j$,
$t_L=t+\gamma$ and $t_R=t-\gamma$ with $\gamma$ characterizing the asymmetric hopping amplitude,
$\phi_h$ corresponds to an applied magnetic flux through a finite ring with length $L$,
and $V_j=V\cos(2\pi \alpha j+\phi_o/L + i h)=V[e^{i(2\pi j\alpha+\phi_o/L)}e^{-h}+e^{-i(2\pi j\alpha+\phi_o/L)}e^{h}]/2$
with $V$, $\phi_o$ and $h$ being real parameters and
$\alpha$ determining the period of the modulation that is taken as an irrational
number in the incommensurate case.
Note that Ref.~\cite{Longhi2019PRL} considered
the case for $\gamma=0$ and $\phi_h=0$. The Hamiltonian can also be written as
$\hat{H}=\hat{\Psi}^\dagger H_r\hat{\Psi}$, where $\hat{\Psi}^\dagger=(\begin{array}{cccc}
                                                                \hat{c}_1^\dagger & \hat{c}_2^\dagger & \cdots & \hat{c}_L^\dagger
                                                              \end{array})
$.

We now write this Hamiltonian in the Fourier space as
\begin{eqnarray}
\hat{H}=&&\sum_{k}[\frac{V}{2}(e^{i\phi_o/L}e^{-h}\hat{a}_{k+1}^\dagger\hat{a}_{k}+e^{-i\phi_o/L}e^h \hat{a}_{k}^\dagger\hat{a}_{k+1}) \nonumber \\
&&+U_k \hat{a}_k^\dagger \hat{a}_k],
\label{Hk}
\end{eqnarray}
where $U_k=2J\cos(2\pi k\alpha+\phi_h/L+ir)=e^{i2\pi k\alpha+\phi_h/L}(t+\gamma)+e^{-i(2\pi k\alpha+\phi_h/L)}(t-\gamma)$ with $J=\sqrt{t^2-\gamma^2}$ and $r=\ln\sqrt{\frac{t-\gamma}{t+\gamma}}$.
The Hamiltonian can be written in a compact form as $\hat{H}=\hat{\Phi}^\dagger H_F\hat{\Phi}$, where $\hat{\Phi}^\dagger=(\begin{array}{cccc}
                                                                \hat{a}_1^\dagger & \hat{a}_2^\dagger & \cdots & \hat{a}_L^\dagger
                                                              \end{array})
$. Evidently, $h$ contributes asymmetric hopping in the Fourier space.

\begin{figure}[t]
  \includegraphics[width=3.4in]{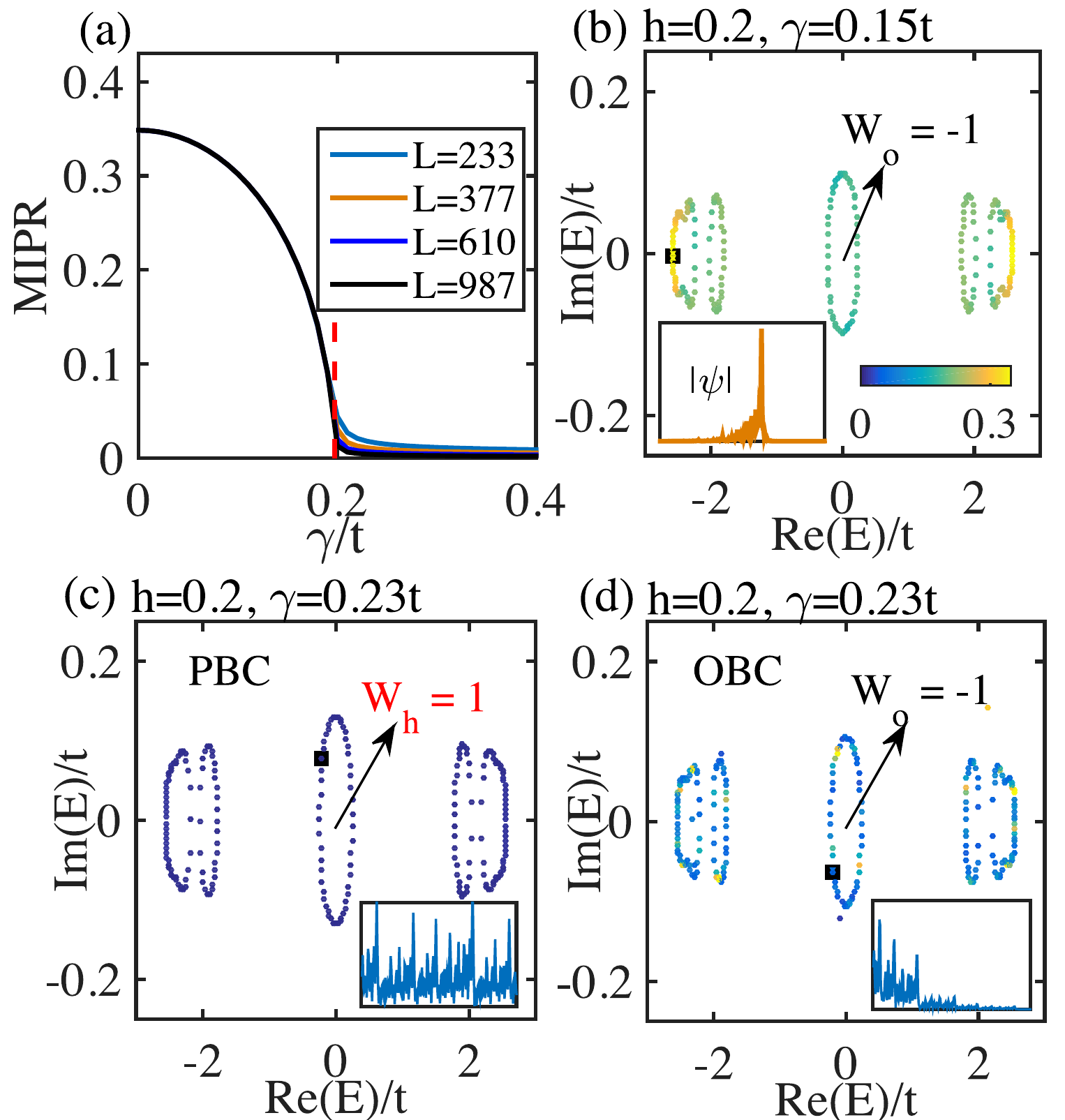}
  \caption{(Color online) (a) MIPR vs $\gamma$ for distinct system sizes. The dashed red line denotes the critical point of
  $\gamma=0.1974t$. Energy spectra in the complex energy plane with the color bar indicating the IPR values of the eigenstates for (b) $\gamma=0.15t$ and (c),(d) $\gamma=0.23t$.
  In (b), the spectra obtained under PBCs are the same as those obtained under OBCs.
  (c) is obtained under PBCs, while (d) is obtained under OBCs.
  The insets plot the amplitudes of the corresponding wave functions as labeled by the black squares.
  The black and red numbers denote the winding numbers of $W_o$ and $W_h$ for the energy loops, respectively.
  Clearly, in (b), the states are localized, while in (c) the states are extended under PBCs, which exhibit non-Hermitian skin effects
  under OBCs.
  Here, $\alpha=(\sqrt{5}-1)/2$, $\phi_o=\phi_h=0$, $L=233$, $h=0.2$, and $V=1.96t$.}
\label{fig1}
\end{figure}

Clearly, when $\phi_h=\phi_o$, $t=V\cosh(h)/2$, and $\gamma=V\sinh(h)/2$,
we have $H_F=H_r^*$ ($*$ denotes the complex conjugate
operation), showing a self-dual symmetry.
Note that the generic case requires $|V\cosh(h)/(2t)|=1$ and $|V\sinh(h)/2|=|\gamma|$.
This symmetry dictates the phase transition between extended and localized
states in terms of $V$, $h$, and $\gamma$.
For instance, when $V=1.96t$ and $h=0.2$, the symmetry gives us the critical point
at $\gamma=0.1974t$, which has been numerically confirmed in Fig.~\ref{fig1}(a) by the change of the
mean inverse participation ratio (MIPR) defined as $I_m\equiv \frac{1}{L}\sum_E \text{I}(E)$. Here, $\text{I}(E)=\sum_j |\psi_j(E)|^4/(\sum_j |\psi_j(E)|^2)^2$
is the inverse participation ratio (IPR) for the right eigenvector
$\psi(E)$ of $H_r$ corresponding to the eigenenergy $E$ with $\psi_j(E)$ representing the
$j$th entry of $\psi(E)$. In the following, we will use the IPR to characterize the localization property
of an eigenstate. Specifically,
for a localized state, the IPR approaches to around $1$, whereas for an extended state, the IPR is of the order of $1/L$.
We note that the self-dual symmetry only determines parts of critical points, i.e., four points 
when $|V/t|<2$ and one point when $|V/t|=2$ in the $(h,\gamma/t)$ plane. The critical lines in the plane can be
approximately determined by $\gamma=V\sinh(h)/2$; see the Appendix.

\begin{figure*}[t]
  \includegraphics[width=\textwidth]{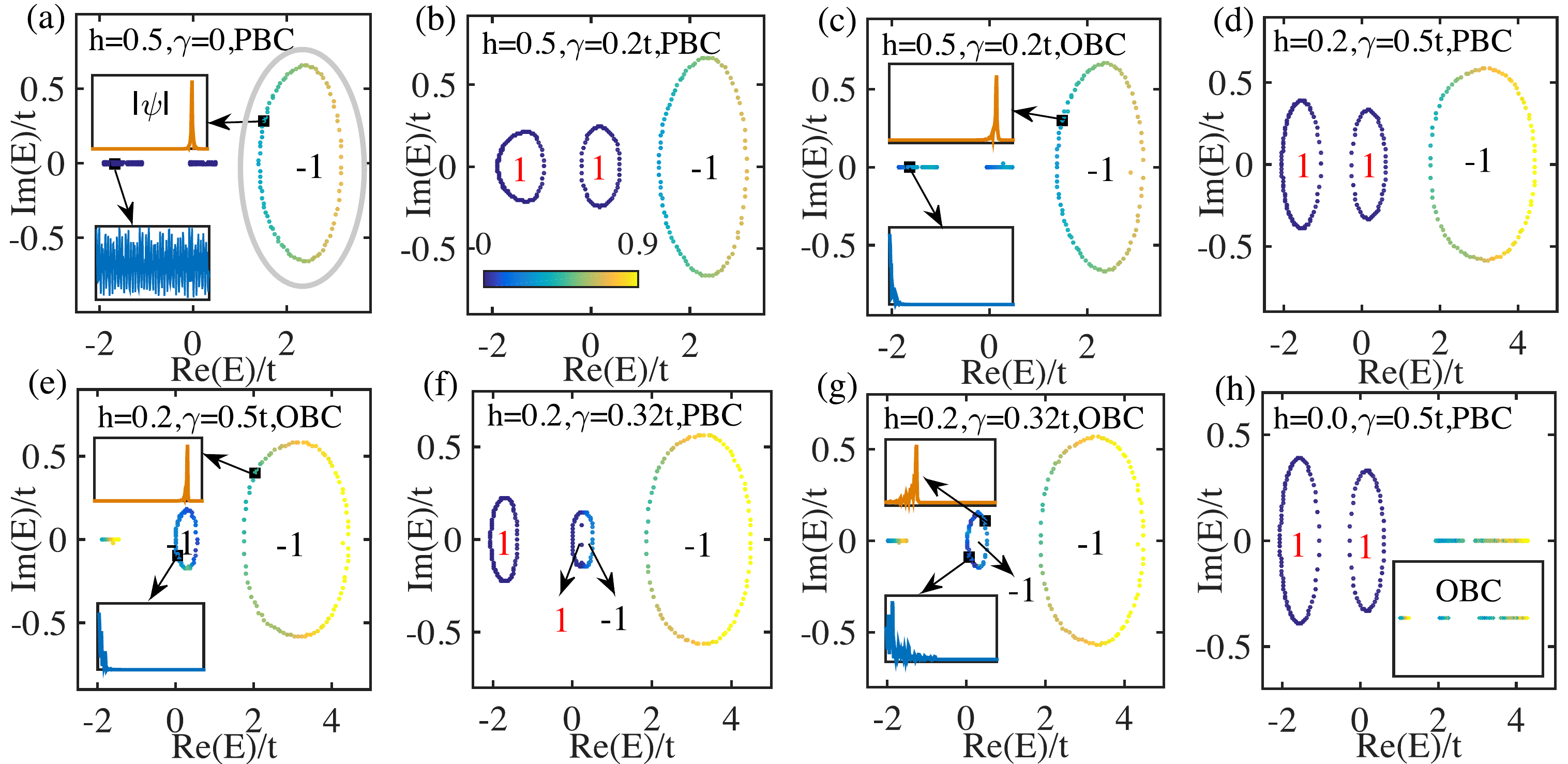}
  \caption{(Color online) Energy spectra in the complex energy plane for systems with (a) $h=0.5$ and $\gamma=0$ under PBCs,
  (b),(c) $h=0.5$ and $\gamma=0.2t$ under PBCs and OBCs, respectively,
  (d),(e) $h=0.2$ and $\gamma=0.5t$ under PBCs and OBCs, respectively, (f),(g) $h=0.2$
  and $\gamma=0.32t$ under PBCs and OBCs, respectively, and (h) $h=0$ and $\gamma=0.5t$ under PBCs (the inset shows the spectrum under OBCs). The color bar indicates the IPR values of the eigenstates.
  The insets plot the amplitude of the wave functions corresponding to the eigenenergy denoted by the black squares.
  The black and red numbers represent the winding numbers of $W_o$ and $W_h$, respectively.
  In (a), the gray loop plots a generalized mobility edge.
  Here we set (a)-(c) $V=0.5t$ and (d)-(h) $V=1.0t$. Other parameters are $\alpha=(\sqrt{5}-1)/2$, $\phi_h=\phi_o=0$, $a=0.5$, and $L=233$.}
\label{fig2}
\end{figure*}

We now define two types of winding numbers measured with respect to a base energy $E_B$ as
\begin{equation}\label{w1}
  \text{W}_\nu(H)=\lim_{L\rightarrow \infty} \frac{1}{2\pi i} \int_{0}^{2\pi} d\phi_\nu \frac{\partial}{\partial \phi_\nu} \ln \det [ H(\phi_\nu) - E_B ],
\end{equation}
with $\nu=h,o$. $W_h(H_r)$ refers to the widely used winding number evaluated by applying a magnetic flux $\phi_h$~\cite{Gong2018PRX}
and $W_o$ refers to the winding number evaluated by applying the phase $\phi_o$
in the onsite potential~\cite{Longhi2019PRL} (i.e., applying a magnetic flux in the Fourier space).
They have been separately utilized to characterize the loop of the energy spectra of extended~\cite{Gong2018PRX} and localized
states~\cite{Longhi2019PRL}, respectively.
It is reasonable as in a periodic boundary geometry, the localized (extended) states are extended (localized) in the Fourier space and thus the winding number is evaluated by
applying a magnetic flux (a phase in the on-site potential) in the Fourier space.
However, under OBCs, only $W_o$ exists since the magnetic flux can only be applied in a ring system.
But it does not mean that only the energy spectra of the localized states can exhibit a loop structure.
In fact, those of extended states can also show similar features.

In the self-duality line for $\phi_h=\phi_o=0$, if $E$ is an eigenenergy of $H_r$, it is also an
eigenenergy of $H_F$ and thus $E^*$ is also an eigenenergy of $H_r$ because $H_r=H_F^*$,
implying that $H_r^*$ and $H_r$ share the same set of eigenvalues.
This equality also gives us $W_o(H_r)=W_h(H_F^*)$.
We thus obtain $W_o(H_r)=-W_h(H_F)=-W_h(H_r)$, indicating that these two winding numbers are either both
nonzero or both zero with respect to $E_B$ under PBCs.
Across the line, if the states are localized, the energy spectra have $W_o=-1$ and $W_h=0$
with the same spectra for PBCs and OBCs, as shown in Fig.~\ref{fig1}(b).
Otherwise, if the states are extended, the energy spectra obtained under PBCs
have $W_h=1$ and $W_o=0$ [see Fig.~\ref{fig1}(c)], suggesting the presence of the non-Hermitian skin effects
under OBCs. But that does not mean that the winding number cannot exist in this scenario.
In fact, we find that the energy spectra obtained under OBCs can still form loops
characterized by $W_o=-1$ despite the presence of the skin effects, as shown in Fig.~\ref{fig1}(d).
We note that this feature exists not only in the incommensurate case,
but also in a commensurate case, which is a periodic system (see Fig.~\ref{fig5} in the Appendix).

\section{Topological mobility edge}\label{sect3}
To generate the mobility edge, we consider the Hamiltonian in~(\ref{HrEq}) with
the onsite potential replaced with
\begin{equation}
  V_j\rightarrow V_j^\prime=\frac{2 V_j}{1-a V_j/V},
\label{TMEModel}
\end{equation}
where $a$ is a real parameter. When $\gamma=h=0$, this model is Hermitian and hosts mobility edges~\cite{Ganeshan2015PRL}.
In the Hermitian case, the mobility edge is defined as the energy that
separates the localized and extended states in the energy spectrum. Yet, in a non-Hermitian case, given that the energies
become complex, we define a generalized mobility edge as boundaries in the complex energy plane that
separate the localized and extended states.

Without $\gamma$, the model has the $\mathcal{PT}$ symmetry, i.e.,
$\mathcal{PT}\hat{H}(\mathcal{PT})^{-1}=\hat{H}$ with
$\mathcal{PT}\hat{c}_j(\mathcal{PT})^{-1}=\hat{c}_{-j}$ and $\mathcal{PT}i(\mathcal{PT})^{-1}=-i$.
Interestingly, we find that the energy spectra obtained under PBCs and OBCs are
identical and parts of them
are real ($\mathcal{PT}$-symmetry preserved) and parts are complex ($\mathcal{PT}$-symmetry broken)
forming a loop,
as shown in Fig.~\ref{fig2}(a). The states with complex energies are
localized and thus the loop is characterized by $W_o=-1$ while those with
real energies are extended with $W_o=W_h=0$. The gray loop in Fig.~\ref{fig2}(a) shows a
generalized mobility edge with localized states inside the loop and extended states outside it; the mobility
edge is clearly not unique. Given that the topological
properties of the energy spectra inside and outside the edge are distinct, we call
it the topological mobility edge.

With $\gamma$ breaking the $\mathcal{PT}$ symmetry,
the energies of the extended states obtained under PBCs
also become complex and form loops, as shown in Figs.~\ref{fig2}(b), ~\ref{fig2}(d) and ~\ref{fig2}(f).
The loops associated with the localized and extended
states are characterized by the winding numbers $W_o=-1$ and $W_h=1$, respectively.
Similarly, we can choose a loop in the complex energy plane as a generalized mobility edge.
The generalized mobility edge is also of a topological nature in the sense that the energy spectra inside it
have $W_o=-1$ and $W_h=0$
and those outside it have $W_h=1$ and $W_o=0$, provided that the base energy is inside the complex energy loop.
In a geometry with open boundaries,
the extended states associated with nonzero $W_h$ are localized at the boundaries
due to the non-Hermitian skin effect, consistent with the prediction in Refs.~\cite{Okuma2020PRL,Zhang2019arxiv},
whereas the localized states are immune to the skin effect. One can also find a topological
mobility edge in this boundary condition.

Although the extended states associated with nonzero $W_h$ under PBCs suffer from the skin effect,
the energy spectra of these states under OBCs can still exhibit topological properties.
For instance, when $h=0.2$ and $\gamma=0.5t$, we observe that the middle loop in the energy spectra
under PBCs becomes a smaller loop under OBCs that are characterized by $W_o=-1$ [see Figs.~\ref{fig2}(d) and ~\ref{fig2}(e)],
instead of a line with zero $W_o$~\cite{Okuma2020PRL,Zhang2019arxiv}. Interestingly,
as we decrease $\gamma$ to $\gamma=0.32t$, we find that
the middle loop in the energy spectra under PBCs deforms into two loops with a partition in the center
[see Fig.~\ref{fig2}(f)]. When the base energy resides inside one (the other) loop,
we have $W_h=1$ and $W_o=0$ ($W_h=0$ and $W_o=-1$). Thus, the states with energies on one (the other) loop
are extended (localized), implying that the common partition is part of the generalized mobility edge.
In this periodic boundary case, the generalized mobility edge is still topological.
But in the open boundary case, these two loops become one, with the energy spectra of the localized states
remaining unchanged, and other states on the loop undergo the skin effect.
This closed loop is also characterized by $W_o=-1$. Evidently, the generalized mobility
edge is not topological in this case because $W_o=-1$ occurs both inside and outside the
closed mobility edge.

In another limit with only asymmetric hopping, the energy spectra of the extended states
obtained under PBCs form closed loops, whereas those of the localized states are real
[see Fig.~\ref{fig2}(h)].
Since the localized states are immune to the skin effects,
their energy spectra obtained under OBCs remain the same as the spectra for PBCs.
But for the extended states, their energy spectra become either real or a smaller loop
due to the skin effects. In this scenario, the generalized mobility
edge is topological in a system with periodic boundaries but not with open
boundaries.

\begin{figure}[t]
  \includegraphics[width=3in]{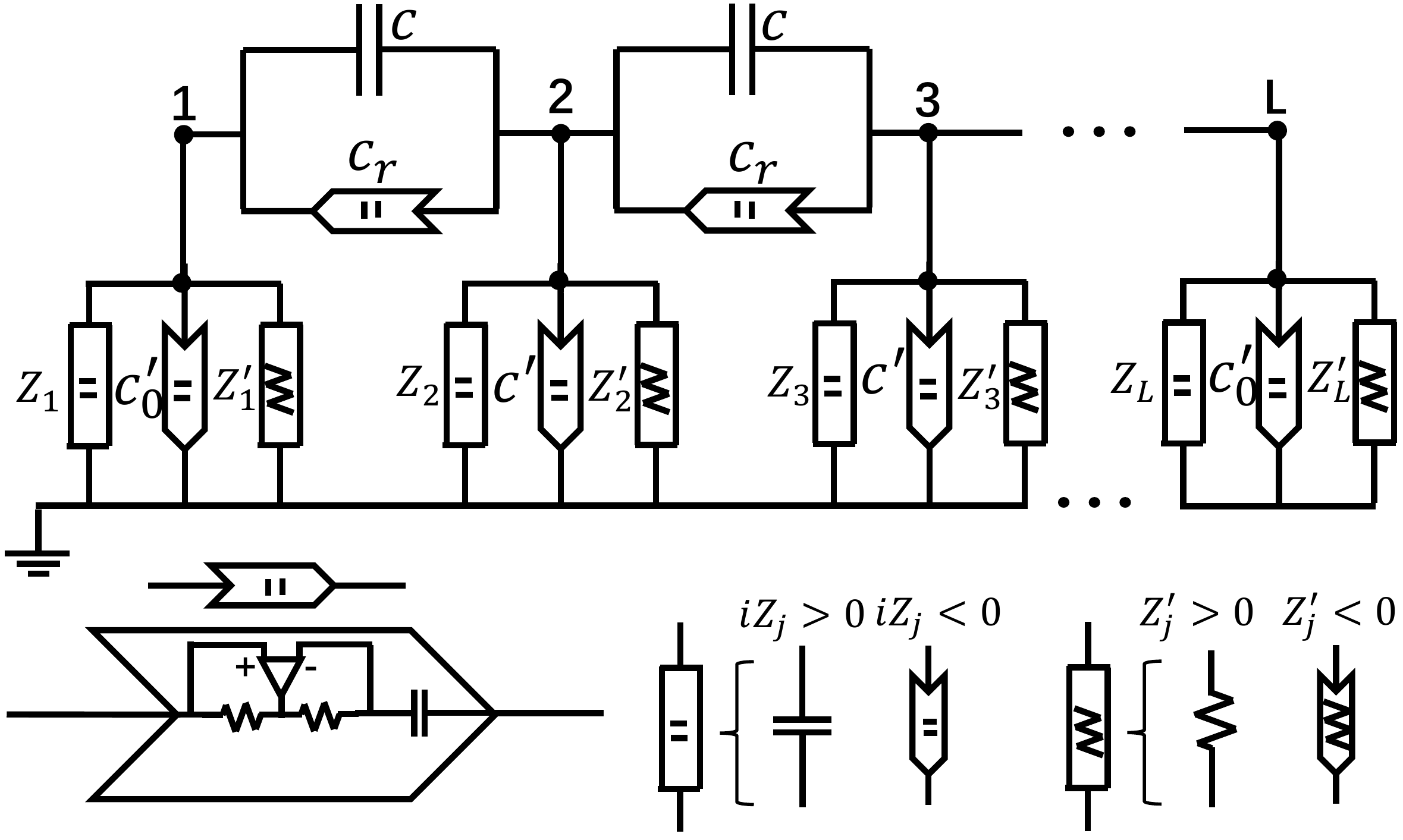}
  \caption{(Color online) The electric circuit for realizing the model Hamiltonian in Eq. (\ref{TMEModel}).
  $C$ and $C_r$ denote the capacitance of the capacitor and INIC for a capacitor, respectively; these
  two electric devices connect one node to its neighboring one, simulating the hopping in the Hamiltonian.
  Here two types of INIC are employed: one for a capacitor (see the bottom left corner for its detailed structure) and the other
  for a resistor, which is produced by replacing the capacitor in the former INIC with a resistor. They are
  represented by the INIC symbol with a capacitor or resistor symbol inside, respectively.
  The on-site potential $V_j^\prime$ at each site is simulated by grounding each node with three suitable devices
  chosen according to the values of their impedance as shown in the bottom right corner.
  To simulate the Hamiltonian, we set $C=t$, $C_r=\gamma$, $C^\prime=2t$, $C_0^\prime=t$, $Z_j=-1/(i\omega\text{Re}(V_j^\prime))$,
  and $Z_j^\prime=-1/(\omega \text{Im}(V_j^\prime))$.}
\label{fig3}
\end{figure}

\section{Experimental realization}\label{sect4}
Here we propose an experimental scheme with electric circuits to simulate the lattice
model, as shown in Fig.~\ref{fig3}. The hopping between neighboring sites is simulated
by capacitors and the negative impedance converter with current inversion (INIC)~\cite{Thomale2019PRL,WKChen},
and the on-site modulations are provided by grounding each node with appropriate electric devices (see Fig.~\ref{fig3}).
After arranging the current and voltage at each node into column vectors $\boldsymbol{I}$ and $\boldsymbol{U}$, respectively, we can write $\boldsymbol{I}=J \boldsymbol{U}$ with $J=-i \omega H_r$ ($\omega$ is the frequency of the current) being the Laplacian of the circuit, which can simulate the Hamiltonian matrix $H_r$.
The energy spectra can be evaluated by measuring the two-point impedances~\cite{Zeng2019arxiv}.
Other platforms, such as the atomic optical lattices~\cite{Ganeshan2015PRL} and photonic systems~\cite{Longhi2019PRL}, are also feasible for the experimental realizations.

\section{Summary}\label{sect5}
In this paper, we report a self-dual symmetry in a non-Hermitian AAH model determining the quantum phase transition
between localized and extended states. We show that there are two types of winding numbers $W_o$ and $W_h$ dictating the topological properties across the transition, i.e., the energies of localized and extended states under PBCs
are characterized by $W_o=-1$ and $W_h=1$, respectively. We further demonstrate the
existence of topological mobility edges. Our work deepens our understanding of the winding numbers and
mobility edges and hence opens the door to further study the generalized mobility edges in non-Hermitian systems.

\emph{Note added.} Recently, we became aware of a related work on mobility edges in non-Hermitian systems~\citep{Chen2020PRB}.

\begin{acknowledgments}
We thank Y.-B. Yang for helpful discussions. This work is supported by the start-up fund from Tsinghua University, the National Thousand-Young-Talents Program and the National Natural Science Foundation of China (Grant No. 11974201).
\end{acknowledgments}

\section*{Appendix}
\begin{figure*}[t]
  \includegraphics[width=\textwidth]{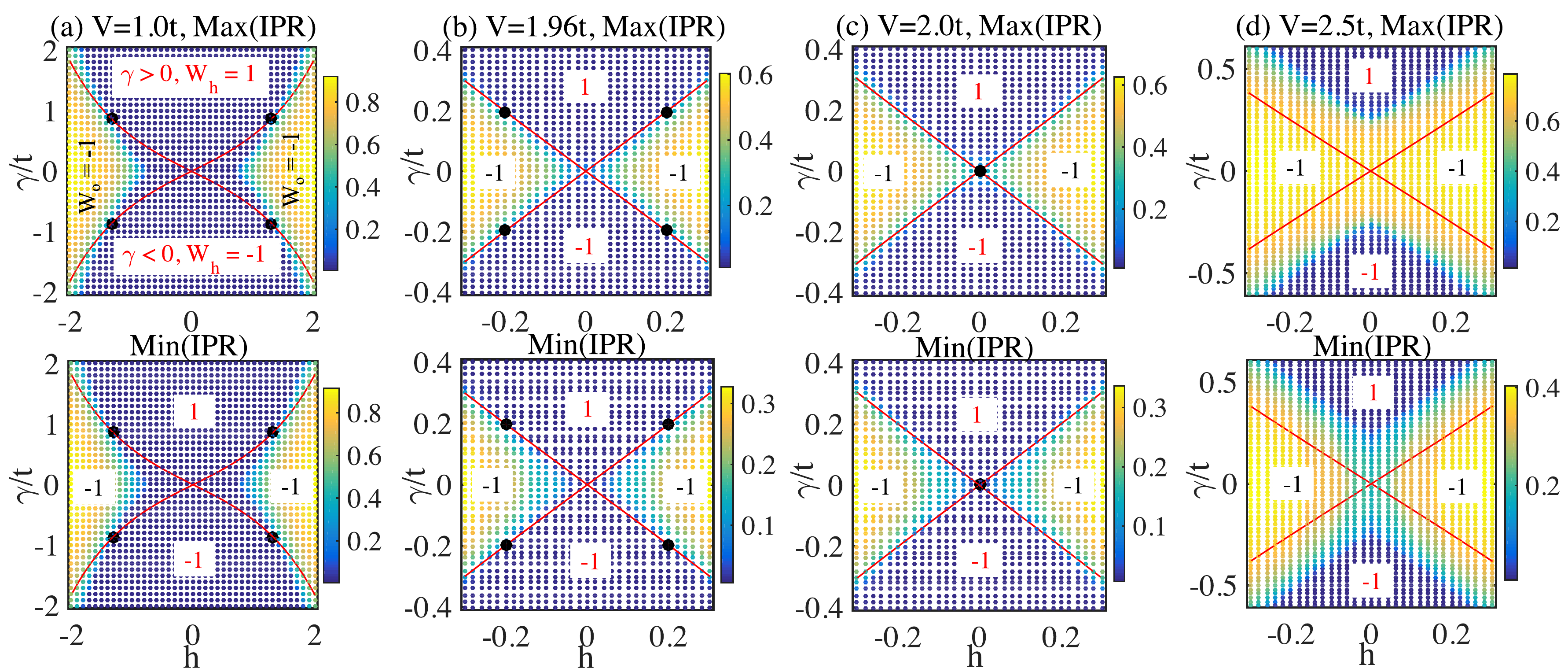}
  \caption{(Color online) The maximum and minimum values of the IPR in the $(h,\gamma/t)$ plane
  for distinct values of $V$: (a) $V=t$, (b) $V=1.96t$, (c) $V=2t$, and $(d)$ $V=2.5t$.
  The IPR is calculated for the right eigenvectors of the Hamiltonian (1) in the main text in a geometry with periodic boundaries.
  The black circles are determined by the self-dual symmetry and the red lines are determined by $\gamma=V\sinh(h)/2$,
  one of the equations from the self-dual symmetry. Here, $\alpha=(\sqrt{5}-1)/2$, $\phi_o=\phi_h=0$,
  and $L=233$.}
\label{fig4}
\end{figure*}

In the Appendix, we first show the maximum and minimum values of the IPR for the eigenstates of 
the Hamiltonian (1) in the main text with respect to $h$ and $\gamma/t$
for different values of $V$ in Fig.~\ref{fig4}. The maximum and minimum values coincide with each other, suggesting the absence of
the mobility edge. The self-dual symmetry determines four critical points when $|V/t|<2$, one when
$|V/t|=2$, and zero when $|V/t|>2$; these points are represented by black circles in the figures. 
We can also see that one of the two equations from the self-dual symmetry, $\gamma=V\sinh(h)/2$,
approximately determines the boundary between the localized and extended states; the approximation works well 
near the critical points evaluated by the self-dual symmetry. Furthermore, the topological properties of the extended 
and localized states can be clearly seen from the winding numbers $W_o$ and $W_h$ labeled in the figure. In the localized region, $W_o=-1$, and
in the extended region, $W_h=\pm 1$, reflecting the topological feature of the quantum phase transition. 
 
\begin{figure*}[t]
  \includegraphics[width=\textwidth]{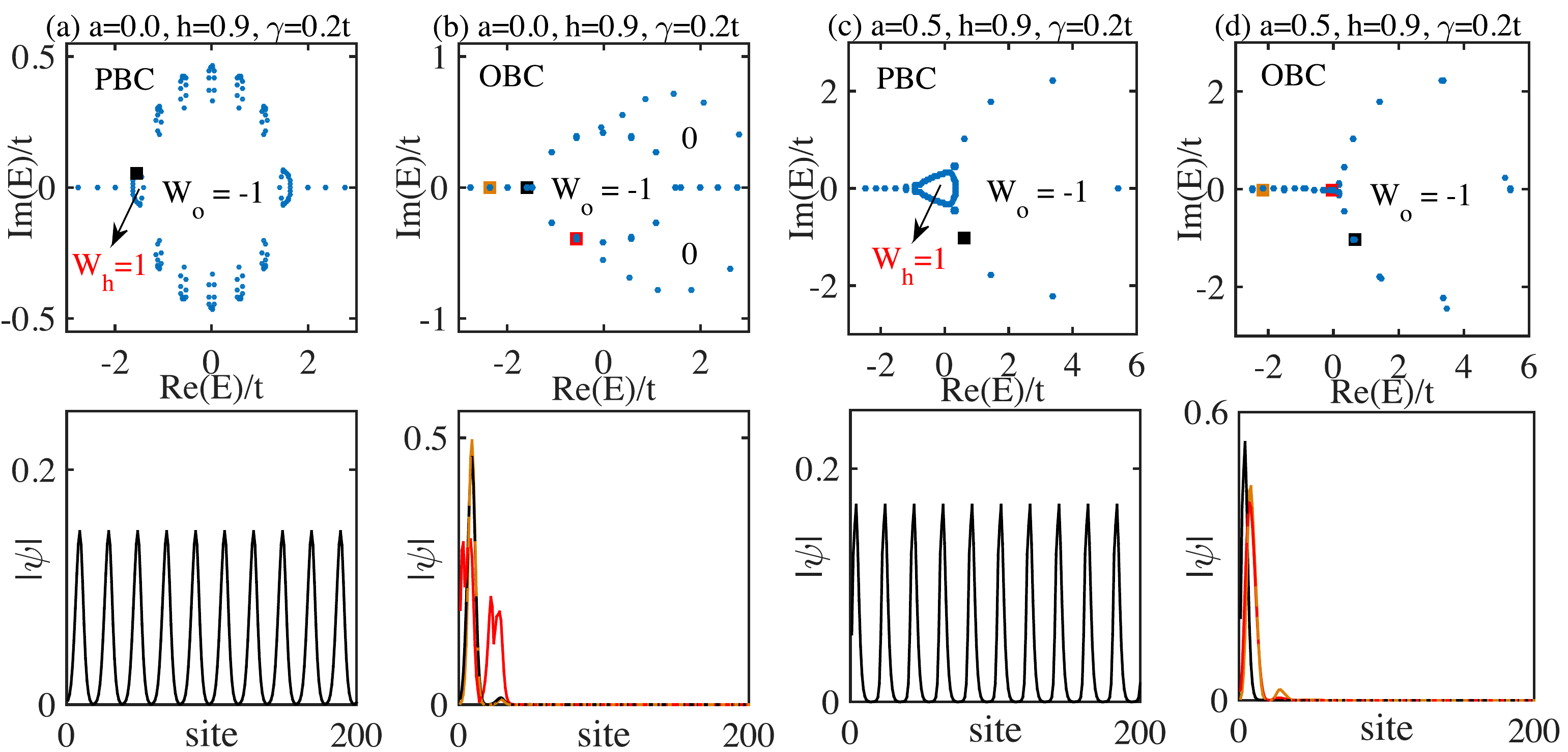}
  \caption{(Color online) Energy spectra in the complex energy plane for a system with commensurate on-site modulations
  with (a)(b) $a=0$ under PBCs and OBCs, respectively, and (c)(d) $a=0.5$ under PBCs and OBCs, respectively.
  The figures in the bottom layer plot the amplitude of the wave functions corresponding to the energies denoted by the squares
  in the corresponding figure above with the same color as that of the wave function.
  The $W_o$ and $W_h$ indicate the winding numbers if the base energy locates inside the loops formed by the eigenenergies of the system. Here, $V=0.5t$, $\alpha=1/20$, $h=0.9$, $\gamma=0.2t$, $\phi_h=\phi_o=0$, and $L=200$.}
\label{fig5}
\end{figure*}

Second, we study the generalized non-Hermitian AAH model with commensurate onsite modulations.
Specifically, when $\alpha=1/20$, for both $a=0$ and $a=0.5$, the energy spectra obtained under PBCs
form loops in the complex energy plane characterized by either $W_h=1$ or $W_o=-1$, as shown in Fig.~\ref{fig5}.
However, under OBCs, we see that the loops with $W_h=1$ disappear, while the others with $W_o=-1$ persist.
Interestingly, all the states including the states with or without winding numbers are localized
at the left boundary due to the non-Hermitian skin effects. This shows that even in a translation invariant
system, the winding number can exist in a system with open boundaries.

\end{document}